\documentclass[10pt]{article}
\usepackage{latexsym}
\usepackage{amsmath}
\begin{document}
\begin{flushright} ULB-TH-00/25 \\  UMH-MG-00/05
  \\ \end{flushright}
\vspace{.1cm}

\begin{center} {\large Uniqueness of the asymptotic $AdS_3$ geometry}\\
\vspace{.5cm} M.~Rooman${}^{a,}$\footnote{
E-mail : mrooman@ulb.ac.be, FNRS senior research associate} 
and Ph.~Spindel${}^{b,}$\footnote{ E-mail :
spindel@umh.ac.be}\\
\vspace{.3cm}  {${}^a${\it Service de Physique Th\'eorique}}\\
{\it Universit\'e
Libre de Bruxelles, Campus Plaine, C.P.225}\\ {\it Boulevard du Triomphe,

B-1050 Bruxelles, Belgium}\\
\vspace{.2cm} {${}^b${\it
M\'ecanique et Gravitation}}\\ {\it Universit\'e de Mons-Hainaut, 20
Place du Parc}\\
{\it 7000 Mons, Belgium}\\ \end{center}
\vspace{.1cm}

\newcommand{\asym}{\mathop {\simeq} \limits_{\bar r \rightarrow \infty}}
\newcommand{\real}{{\hbox{{\rm I}\kern-.2em\hbox{\rm R}}}}

\begin{abstract}
We explicitly show that in (2+1) dimensions the general solution 
of the Einstein equations with negative cosmological constant on a 
neigbourhood of timelike spatial infinity can be obtained from BTZ metrics 
by coordinate transformations corresponding geometrically to 
deformations of their spatial infinity surface. Thus, whatever the topology
and geometry of the bulk, the metric on the timelike extremities is BTZ.
\end{abstract}

Graham and Lee \cite{GL} proved that, under suitable topological
assumptions, Euclidean
Einstein spaces with negative cosmological constant $\Lambda$
are completely defined by the geometry on
their boundary. Furthermore, Fefferman and Graham (FG) \cite{FG}
showed that, whatever the signature,
there exists an asymptotic expansion of the metric, which formally
solves the Einstein equations with $\Lambda=-1/\ell^2<0$;
we choose hereafter the length units such that $\ell=1$.
The first terms of this expansion may be given by even powers of
a radial coordinate $r$~:
\begin{equation}
\label{MFG}
ds^2
\asym
  {dr^2\over{r^2}} \, + \,
{r^2} \stackrel {(0)} {\bf g} (x^i)  \, + \,
\stackrel {(2)} {\bf g} (x^i) \, + \, \cdots \qquad .
\end{equation}
On $(n+1)$-dimensional space-times, the full asymptotic expansion
continues with terms of negative even  powers of $r$ up
to $r^{-2([\frac {n+1} 2]-2)}$, with in addition a logarithmic term of
the order
of $r^{-(n-2)}\log r$ when $n$ is even and larger than 2. All these terms
are completely defined by the boundary geometry
$\stackrel {(0)} {\bf g} (x^i)$, assumed to be non-degenerate
(i.e. n-dimensional and Lorentzian).
They are
followed by terms of all negative powers
starting from $r^{-(n-2)}$. The trace-free part of the $r^{-(n-2)}$
coefficient is not fully determined by $\stackrel {(0)} {\bf g}$; it
contains degrees of freedom in Lorentzian spaces \cite{NB,BERS}.
Once this ambiguity is fixed, all subsequent terms become determined.

On (2+1)-dimensional spaces, Ba\~nados \cite{B}
showed, for flat boundary geometries, that the FG expansion stops at
order $r^{-2}$ on a neighbourhood 
${\cal V}_\infty\simeq{\cal I} \times \Sigma_\infty$,
where $\Sigma_\infty$ is a timelike component of the surface at
infinity and $\cal I$ is a semi-infinite open interval $]r_0,\infty[$ of 
the $r$-variable;
such a neighbourhood is hereafter called an extremity.
The most general asymptotic solution\footnote{By asymptotic solution 
we mean an exact solution of the Einstein
  equations with negative cosmological constant, which is valid on 
${\cal V}_\infty$, but which does not necessarily admit a singularity-free
extension.}
he obtains is of the form~:
\begin{equation}
\label{MBan}
ds^2=\frac{ dr^2} {r^2} +
(r  dx^+ + \frac 1 r {\cal L}_-(x^-) dx^-)
(r  dx^- + \frac 1 r {\cal L}_+(x^+) dx^+) \qquad .
\end{equation}
This metric depends on two functions of one variable,
${\cal L}_+(x^+)$ and ${\cal L}_-(x^-)$. Skenderis and Solodukhin
\cite{SS} generalized this result by showing that in (n+1) dimensions,
for any locally AdS metric, i.e. any conformally flat metric with
Riemann tensor
$R^{\mu \nu}_{\ \ \rho \sigma}= 
   \delta ^\mu _\sigma \delta ^\nu _\rho - 
   \delta ^\mu _\rho \delta ^\nu _\sigma$,
the FG expansion can be written as~:
\begin{equation}
\label{MSS}
ds^2= \frac{dr^2}{r^2} + 
\left ( r \delta ^k _i + \frac 1 {2r}\stackrel {(2)} g _{il} 
    \stackrel {(0)} g \strut ^{lk} \right )
\stackrel {(0)} g _{km}
\left ( r \delta ^m _j + \frac 1 {2r} \stackrel {(0)} g \strut ^{mp}
  \stackrel {(2)} g _{pj} \right ) dx^i dx^j \quad ,
\end{equation}
where $\stackrel {(0)} {\bf g}$ is a conformally flat boundary metric
\cite{RS}.

In this letter, we explicit the link between the $AdS_3$ metrics 
(\ref{MBan}, \ref{MSS}),
the BTZ black hole metrics \cite{BTZ} characterized by their mass $M$ and
angular momentum $J$, and the Liouville fields
\cite{CHD, RS, SS} that encode the boundary degrees of freedom, under the 
assumption that the surface at infinity is topologically a cylinder~:
$\Sigma_\infty\simeq\real \times S^1$. 
Since solutions with different values of $M$ and $J$ are distinct, 
the question arises whether or not solutions in which the constants 
$M$ and $J$ are
replaced by functions are more general. 
Our starting
point is the observation that the coordinates $(r, x^i)$ constitute
a Gaussian coordinate system (see ref. \cite{L}, chap. 3)
with respect to the family of surfaces
$r=cst$; the boundary corresponds to $r=\infty$.
Accordingly, we shall examine Gaussian coordinate systems built on families of
surfaces, embedded in BTZ spaces, which asymptotically merge into the
boundaries of these spaces.

The BTZ black hole geometry is characterized by two parameters,
the mass $M$ and the
angular momentum $J$;
for simplicity we restrict ourselves first to the non-extremal
case $M>|J|$.
Its metric reads as~:
\begin{equation}
\label{MBTZ}
ds^2 = -(\rho^2 -M) dt^2 +
\frac {d\rho ^2} {\rho^2 -M + \frac {J^2}{ 4 \rho^2}}
+ \rho^2 d\varphi^2 + J d\varphi dt \qquad .
\end{equation}
Posing $M_\pm = \frac 1 4 (M \pm  J )$ and performing the
coordinate transformation~:
\begin{eqnarray}
\xi^2 &=& \frac { \rho^2 -M_+ -M_- +\rho
  \sqrt{ \rho^2 -2(M_+ + M_-)+
\frac 1  {\rho^2}(M_+ - M_-)^2} } {2 \sqrt{M_+ M_-}} \quad ,\\
y^\pm &=& \sqrt{M_\pm} \, (\varphi \pm t)  \qquad ,
\end{eqnarray}
this metric can be written in the FG form~:
\begin{equation}
\label{MFGBTZ}
ds^2= \frac{ d\xi^2} {\xi^2} +
(\xi  dy^+ + \frac 1 \xi dy^-)
(\xi  dy^- + \frac 1 \xi dy^+) \qquad .
\end{equation}
This form reflects the local equivalence of all the BTZ metrics. Their
global differences are encoded in the periods of the variables, with
the identification $y^\pm+2\pi\sqrt{M_\pm} \equiv y^\pm$.

The metric (\ref{MFGBTZ}) admits two Killing vectors $\partial_{y^\pm}$.
This allows to integrate by quadratures the geodesic
equations in these coordinates. Denoting by $C_\pm$ the (constant)
values of the moments conjugate to $y^\pm $, by $\gamma$ an affine
parameter on these geodesics, and by $\gamma_0$ and $z^\pm$
integration constants, the spacelike geodesics can be written as~:
\begin{eqnarray}
\label{transfo1}
\gamma - \gamma_0 &=& \log [\, \Gamma (\xi,C_+,C_-)]-
  \log[\, \Gamma(\xi_0,C_+,C_-)]
\qquad ,\\
\label{transfo2}
y^\pm - z^\pm &=& Y^\pm(\xi,C_+,C_-)- Y^\pm(\xi_0,C_+,C_-)  \qquad ,
\end{eqnarray}
where we have introduced the primitives expressed in terms of
  $X=\xi^2 + \xi^{-2}$~:
\begin{eqnarray}
\label{intg}
&& \log [ \, \Gamma(\xi,C_+,C_-)] =
\int \frac {\xi^4-1}
   {\sqrt{ (\xi^4-1)^2- 4 \xi^2 (C_- \xi^2 -C_+)(C_+ \xi^2-C_-)}} \,
   \frac {d\xi} \xi \nonumber \\
&& \qquad = \frac 12 \log \frac 12 [X-2 C_+C_- +
   \sqrt{X^2-4C_+C_-X+4 (C_+^2+C_-^2-1)}]
    \, ,
\end{eqnarray}
\begin{small}
\begin{eqnarray}
\label{inty}
&& Y^\pm(\xi,C_+,C_-) =
  2 \int
  \frac {\xi (C_\mp -2 C_\pm \xi^2 + C_\mp \xi^4)\qquad \qquad d\xi}
  {(\xi^4-1) \sqrt{ (\xi^4-1)^2- 4 \xi^2 (C_- \xi^2 -C_+)(C_+ \xi^2-C_-)}}
   \nonumber \\
&& \quad = \frac 14  \left  \{ \vrule height 7mm width 0mm
   \log \frac{(2+X)\,( C_+C_- -C_+-C_-+1) }{(2-X)\,(C_+C_- \mp C_+  \pm C_- -1)}
   \right . \\
&& \,
     \left . +\log \frac {2(C_+C_--C_+^2-C_-^2+1)-
          (1-C_+C_-)X\pm(C_--C_+)\sqrt{X^2-4C_+C_-X+4
  (C_+^2+C_-^2-1)}}
  {2(C_+C_-+C_+^2+C_-^2-1)-
   (1+C_+C_-)X+(C_-+C_+)\sqrt{X^2-4C_+C_-X+4 (C_+^2+C_-^2-1)}}
  \vrule height 7mm width 0mm \right \}
   \, . \nonumber
\end{eqnarray}
\end{small}

Consider now the family of surfaces $\Sigma(\rho_0)$~:
\begin{equation}
\xi e^{-\Delta(y^+,y^-)} = \rho_0 \qquad ,
\end{equation}
where the function $\Delta$ has the periodicities $2\pi \sqrt{M_\pm}$
of $y^+$ and $y^-$.
In the limit where $\rho_0$
tends to $\infty$, the geodesic congruence, orthogonal to $\Sigma(\rho_0)$ at
the point of coordinates 
$y^\pm=z^\pm$, $\xi=\xi_0\equiv \rho_0 \exp{[\Delta(z^+,z^-)]}$,
is obtained by fixing the constants $C_+$ and $C_-$ as~:
\begin{equation}
C_\pm= \partial_{y^\pm} \Delta \, |_{y^\pm=z^\pm}
  \equiv {\cal C}_\pm(z^+,z^-) \qquad   .
\end{equation}
If we moreover impose, for all the geodesics of
this congruence, the value $\gamma_0$ of $\gamma$ on  $\Sigma(\rho_0)$ to
be equal to $\log\rho_0$, we obtain in the same limit $\rho_0 
\rightarrow \infty$ a
finite
expression of $\gamma$ as a function of $\xi$ that reads as~:
\begin{equation}
\gamma= \Delta(z^+,z^-) + \log[\, \Gamma(\xi,{\cal C}_+,{\cal C}_-)]  \qquad ,
\end{equation}
which asymptotically leads to~:
\begin{equation}
\gamma= \Delta(z^+,z^-) + \log \xi -{\cal C}_+\,{\cal C}_-\, \xi^{-2}+
\frac 12 ({\cal C}_-^2+{\cal C}_+^2 - 3\,
{\cal C}_-^2\,{\cal C}_+^2)\, \xi^{-4}+ O(\xi^{-6})\qquad .
\end{equation}
Posing $r=\exp (\gamma)$
and defining $\Gamma_*$ as the inverse function of $\Gamma$ and $Z^\pm$ as the
composed functions
$Y^\pm[\Gamma_*(r\, e^{\Delta}, {\cal C}_+, {\cal C}_-),{\cal 
C}_+,{\cal C}_-]$,
we obtain~:
\begin{eqnarray}
\xi&=& \Gamma_*[r \,e^{\Delta},{\cal C}_+,{\cal C}_-]\\
& \simeq& r \, e^{\Delta}\left ( 1+{\cal C}_+{\cal C}_-\,\frac1{(r\/e^{\Delta})^2}-
\frac 12 ({\cal C}_+^2+{\cal C}_-^2) \frac 1{(r\/e^{\Delta})^4} \right ) + O(r^{-5})
\label{approxi}\\
y^\pm&=&z^\pm + Z^\pm[r\/ e^{\Delta},{\cal C}_+,{\cal C}_-]\\
  &\simeq& z^\pm -
  {\cal C}_\mp \frac1{(r\/e^{\Delta})^2}+{\cal C}_\pm (1+{\cal C}_\mp^2)
\frac1{(r\/e^{\Delta})^4} +
O(r^{-6})\qquad ,\label{approy}
\end{eqnarray}
In agreement
with the general statement of \cite{SS}, 
the FG expansion stops at order $r^{-2}$,
whatever the (conformally flat) boundary geometry is. Accordingly, to obtain 
the expression of the metric 
in terms of the new variables $r$ and $z^\pm$,
it suffices to use eqs (\ref{approxi}, \ref{approy}); contributions 
of subsequent terms in the asymptotic
expansions of $\xi$ and $y^\pm$ must indeed cancel out.
Hence, the metric (\ref{MFGBTZ}) becomes~:
\begin{eqnarray}
ds^2 &=& \frac{dr^2}{r^2}+
[1+\frac{e^{-2\Delta}}{r^{2}}\partial^2_{z^+z^-}\Delta][1-(\partial_{z^+}\Delta)^2+
\partial^2_{z^+}\Delta] \,
(dz^+)^2  \nonumber \\
&+&[1+\frac{e^{-2\Delta}}{r^{2}}\partial^2_{z^+z^-}\Delta][1-(\partial_{z^-}\Delta)^2+
\partial^2_{z^-}\Delta] \, (dz^-)^2 \nonumber\\
&+& \left \{ r^2 e^{2\Delta}+ 2 \partial^2_{z^+z^-}\Delta
     \right . \nonumber \\
&+& \left . \frac{e^{-2\Delta}}{r^2} \left [
(\partial^2_{z^+z^-}\Delta)^2+[1-(\partial_{z^+}\Delta)^2+
\partial^2_{z^+}\Delta][1-(\partial_{z^-}\Delta)^2+
\partial^2_{z^-}\Delta] \right ] \right \} dz^+ dz^- \qquad .
\end{eqnarray}
We would like to stress that this expression is not a truncated asymptotic
approximation, but an
exact solution of the Einstein equations, valid in a
neighbourhood of the surface at infinity (a tedious calculation confirms that
the expected cancellations  occur and that
the Ricci tensor is~: $R_\mu^\nu=-2\delta_\mu^\nu$). 

Now, let us split the deformation function as~:
\begin{equation}
\Delta(z^+,z^-)={\cal D}(z^+,z^-) +
\frac 1 2 ( {\cal A}_+(z^+)+ {\cal A}_-(z^-)) \qquad ;
\end{equation}
this form becomes unambiguous once we fix for instance
${\cal D}(z^+,0)=0$ and ${\cal D}(0,z^-)=0$.
We then perform the changes of variables~:
\begin{equation}
\label{xz}
x^\pm=\int e^{{\cal A}_\pm(z^\pm)}dz^\pm \qquad ,
\end{equation}
and rewrite the metric as~:
\begin{eqnarray}
\label{genmet}
ds^2 &=& \frac{dr^2}{r^2}+ 
\left [ r e^D \, dx^+ + 
\frac 1 {r e^D} \left ( L_- \, dx^- + ( \partial^2_{x^+x^-} D) dx^+ \right ) \right ]
\nonumber \\
&& \qquad \left [ r e^D \, dx^- + 
\frac 1 {r e^D} \left ( L_+ \, dx^+ + ( \partial^2_{x^+x^-} D) dx^- \right ) \right ]
\qquad ,
\end{eqnarray}
where~:
\begin{equation}
L_\pm (x^+,x^-)= e^{-2A_\pm}- (\partial_{x^\pm}D)^2
+ \partial^2_{x^\pm} D
+ \frac 14 (\partial_{x^\pm}A_\pm)^2
+ \frac 12 \partial^2_{x^\pm} A_\pm
\qquad ,
\end{equation}
and
$D(x^+,x^-)={\cal D}[z^+(x^+),z^-(x^-)]$,
$A_\pm(x^\pm)={\cal A}_\pm[z^\pm(x^\pm)]$.

Considering the terms of order $r^2$ and $r^0$ in the metric (\ref{genmet}),
we recognize the equation obtained in \cite{BERS,RS,SS,HSS}~:
\begin{equation}
\stackrel {(2)} {g} _{\mu\nu}=\frac12 [ T_{\mu\nu}(\phi^L) -
\stackrel {(0)}{g}_{\mu\nu}\cal{R} ] \qquad ,
\end{equation}
where $\cal R$ is the 2-dimensional Gauss curvature of the asymptotic geometry \\
$\stackrel {(0)}{\bf g}=e^{2 D(x^+,x^-)} dx^+ dx^-$, and
$T_{\mu\nu}[\phi^L]$
the energy-momentum tensor of the Liouville field defined as~:
\begin{equation}
\phi^L(x^+,x^-)=\phi^L_0 (x^+,x^-) -2 D (x^+,x^-) \qquad ,
\end{equation}
where $\phi^L_0$ is directly obtained from
the Liouville field associated to the
BTZ metric (\ref{MFGBTZ}) re-expressed in $x^\pm$ coordinates~:
\begin{equation}
\label{Liou0}
\phi^L_0 (x^+,x^-)= \log
\frac {| \partial_{x^+} f_+ \, \partial_{x^-} f_-|}{(f_+ +f_-)^2}
\qquad {\rm with}  \qquad
f_\pm(x^\pm)=\frac{ a_\pm e^{2 z^\pm (x^\pm)} + b_\pm}
   { c_\pm e^{2 z^\pm (x^\pm)} + d_\pm}
\end{equation}
depending on the parameters $a_\pm, b_\pm, c_\pm, d_\pm$ such that
$a_\pm\/d_\pm-b_\pm\/c_\pm=1$ (see ref. \cite{RS}).

To be well-defined, the functions ${\cal D} (z^+,z^-)$ and
${{\cal A}_\pm} (z^\pm)$ must be periodic in $z^+$ and $z^-$~:
\begin{eqnarray}
{\cal D}(z^+ +2\pi \sqrt{M_+},z^- +2\pi \sqrt{M_-})&=&{\cal D}(z^+,z^-)
\qquad , \\
{\cal A}_\pm(z^\pm +2\pi \sqrt{M_\pm})&=&{\cal A}_\pm(z^\pm) \qquad .
\end{eqnarray}
Hence, we immediately see that the periods $P_{\pm}$ of the variables
$x^\pm$ are given by~:
\begin{equation}
P_{\pm}= \int_0^{2 \pi \sqrt{M_\pm}} e^{{\cal A}_\pm(z^\pm)} \, dz^\pm \qquad ;
\end{equation}
they are, in general, different from those of the initial variables
$y^\pm$. Thus, if $M$ and $J$ are given, this equation yields directly 
the periodicity of the variables $x^\pm$ and of the functions appearing in
eq. (\ref{genmet}). Conversely, if we start from the expression (\ref{genmet})
of the metric, the functions $D$ and $L_\pm$ allow to determine the
periodicities $P_\pm$ of the $x^\pm$ variables and, as a consequence,
the values of $M$ and $J$ via de equation~:
\begin{equation}
\label{periods}
 2 \pi \sqrt{M_\pm} = \int_0^{P_\pm} e^{-A_\pm(x^\pm)} \, dx^\pm \qquad .
\end{equation}
These considerations clarify the previously noted \cite{RS}
relationship between the eigenvalues of
Wilson loop matrices, the Floquet exponents occurring in the
solutions of the parallel
transport equations, and the mass and angular momentum of the BTZ
black holes, in agreement with the fact that
eigenvalues of holonomy matrices corresponding
to loops of constant
$\xi$ represent  an invariant concept with respect to coordinate
transformations.

Let us conclude with a few remarks. First
we would like to stress that
the metric (\ref{genmet}), obtained from the BTZ metric (\ref{MFGBTZ})
by a change of coordinates, corresponds to the most general form of the
FG asymptotic expansion (\ref{MFG}) in (2+1) dimensions. Indeed,
as all two dimensional geometries with cylindrical topology
are conformally flat, any
boundary geometry can be written as
  $\stackrel {(0)}{\bf g}=e^{2 D(x^+,x^-)} dx^+ dx^-$. Moreover,
$\stackrel {(2)}{\bf g}=L_+ (dx^+)^2 + L_- (dx^-)^2 +
2 (\partial^2_{x^+x^-} D)  dx^+ dx^-$ constitutes
the most general subdominant metric satisfying the Einstein equations and,
according to the FG theorem,
all subsequent terms $\stackrel {(2n)}{\bf g}$ with
$n \ge 2$ are unambiguously fixed by
$\stackrel {(0)}{\bf g}$ and $\stackrel {(2)}{\bf g}$.
This provides an alternative proof in (2+1) dimensions 
that the most general metric expansion stops at order
$r^{-2}$ \cite{SS}.
The metric (\ref{MBan})
is recovered from this general metric  by imposing
a flat boundary geometry.

We showed hereabove the construction of the
FG expansion, starting
from non-extremal BTZ black hole metrics (\ref{MFGBTZ}).
To consider more general cases, it is sufficient to
start from the following generalization of (\ref{MFGBTZ})~:
\begin{equation}
\label{MFGEPS}
ds^2= \frac{ d\xi^2} {\xi^2} +
(\xi  dy^+ + \frac {\epsilon_-} \xi dy^-)
(\xi  dy^- + \frac {\epsilon_+} \xi dy^+) \qquad ,
\end{equation}
where the $\epsilon_\pm$ are the ``signs of $M_\pm$''. This means that  for
non-extremal BTZ black hole metrics $(\epsilon_+,\epsilon_-) =
(1,1)$, for extremal ones $(\epsilon_+,\epsilon_-)$ is
$(1,0)$ or $(0,1)$ when $M=|J|\neq 0$ and $(0,0)$ if $M=|J|=0$;
$(\epsilon_+,\epsilon_-) = (1,-1)$ [resp. $(-1,1)$] with $y^-$ [resp. $y^+$]
non-periodic corresponds
to self [resp. anti self]-dual solutions \cite{CH},
$(\epsilon_+,\epsilon_-) = (-1,-1)$ with both $y^\pm$ non-periodic 
to pure AdS geometry, and the
remaining possibilities with periodic $y^+$ and/or $y^-$ 
to spaces with closed time lines.
For all these cases, we obtain the same final expression  (\ref{genmet})
of the metric, but with the functions $L_\pm$ now defined as~:
\begin{eqnarray}
\label{GenL}
L_\pm (x^+,x^-)= \epsilon_\pm e^{-2A_\pm}- (\partial_{x^\pm}D)^2
+ \partial^2_{x^\pm} D
+ \frac 14 (\partial_{x^\pm}A_\pm)^2
+ \frac 12 \partial^2_{x^\pm} A_\pm
\qquad .
\end{eqnarray}

Furthermore, given an FG expression (\ref{genmet}) of the metric,
the functions
${A}_\pm(x^\pm)$ can be computed  from $D(x^+,x^-)$ and $L_\pm(x^+,x^-)$
using eq.(\ref{GenL}).
If the so obtained functions $\exp{[{A}_\pm(x^\pm)]}$ are positive,
the change of variables (\ref{xz})
is well-defined and the
metric's singularities are just coordinate singularities corresponding
to the locus of points where neighbouring geodesics intersect. If
the functions ${A}_\pm(x^\pm)$ are moreover periodic, the  starting FG metric
is simply a
local description of some asymptotic
BTZ geometry, corresponding to a deformation of the surface at infinity
defined by the standard coordinates $\rho,\ \varphi$ and $t$; the
values of the mass and angular momentum are encoded in the periodicity of
the $z^\pm$ variables, or in a more hidden way in the periodicity of
the $x^\pm$ coordinates via the functions $D(x^+, x^-)$ and  
$L_\pm(x^+, x^-)$. On the other hand,
if the functions ${A}_\pm(x^\pm)$ are not periodic (i.e. for $P_\pm=0$), 
of course no
identifications are allowed and the metric is nothing else than a local metric
on a pure AdS space \cite{BTZ}.
However, we may also consider metrics (\ref{genmet}) constructed
with functions $L_\pm$ that cannot be obtained from (\ref{MFGBTZ})
or (\ref{MFGEPS})
by a change of coordinates, in particular when $A_\pm$ becomes singular and
$e^{-A_\pm}$ negative. In such case, the metric (\ref{genmet}) still
solves the Einstein equations, but may present true singularities, whose
interpretation requires further investigation. They may for instance be the
sign of the presence of multi-black holes in the bulk of the 
space \cite{MS, Br}.

Let us emphasize the significance of our results.
Of course, for fixed $\Lambda$,
all $AdS_3$ metrics are locally isometric, whatever the arbitrary 
$x^\pm$-dependent functions occurring in them (see eq. (\ref {MFGEPS})). 
What we show here is that for any such functions the metric can be transformed
on a whole extremity ${\cal V}_\infty$ of the space into a canonical 
BTZ expression with specified values of $M$ and $J$, via eq. (\ref{periods}).
The degrees of freedom that give rise to the black
hole entropy are thus not more encoded in these functions than in
$M$ and $J$. It is however still conceivable that they are 
encoded through the existence of more complex topologies, with several 
disconnected components of the spatial infinity surface. Finally note that
the coordinate transformations (\ref{transfo1}, \ref{transfo2}) 
are nothing else than the finite
group transformations whose infinitesimal generators provide the well
known Virasoro algebra of asymptotic symmetries \cite{BH}.

{\bf Acknowledgments} We thank S. de Haro for bringing refs \cite{SS, HSS}
to our attention.

\end{document}